\documentclass[a4paper,12pt]{article} 
\usepackage{amsmath}
\usepackage{color}
\usepackage{amssymb, amsfonts, amsthm, amscd}
\usepackage{bm}
\usepackage{graphicx}
\usepackage[all]{xy}

\begin{document}

\vspace*{-15mm}
\begin{flushright}
{\bf OCHA-PP-376} \\
\end{flushright}


\vspace{7mm}

\begin{center}
{\Large\bf
Dyonium Induced Fermion Number Violation}

\vspace{7mm}

{\bf Akio Sugamoto}
\vspace{2mm}

{\it Department of Physics, Graduate School of Humanities and Sciences, Ochanomizu University, 2-1-1 Ohtsuka, Bunkyo-ku, Tokyo 112-8610, Japan}

\vspace{2mm}


\end{center}

\vspace{5mm}
\begin{center}
\begin{minipage}{14cm}
\baselineskip 16pt
\noindent
\begin{abstract}
Dyonium induced fermion number violation is studied in a $SU(2)_L \times U(1)_Y$ gauge theory with a doublet Higgs field.  Dyonium is a generalization of Nambu's monopolium, having finite sized pair of dyon and anti-dyon, connected with a thin string under the linear plus Coulomb force.\footnote{~Monopole has a magnetic charge without electric charge, while dyon has both charges. Monopolium is a bound state of monopole and anti-monopole, while dyonium is that of dyon and anti-dyon.}

This is a follow-up of the author's paper on monopolium induced neutrino mass (including the lepton and baryon number violations), studied with a SO(10) Grand Unified model in 1983.  To fulfill the requirement from the chiral anomaly or its index theorem, an electric field should be excited in parallel to the dipole magnetic field. This crucial dynamical problem, not fully answered then, has been solved, by considering not monopolium but dyonium.

The fermionic zero modes in the dyonium background fields are necessary in evaluating the transition rate of the fermion number violation processes.  Their Dirac equation can be reduced to a single component partial differential equation, similar to the renormalization group equation, which may be useful to estimate the reaction rates.
 \end{abstract}

\end{minipage}
\end{center}

\vspace{5mm}
\section{Introduction}

The important issues to be clarified in the present high energy physics are the baryogenesis and the neutrino mass.  Regarding the baryogenesis, many scenarios have been proposed so far, which satisfy the Sakharov's three criterions \cite{Baryogenesis}.  The popular scenario is the ``sphaleron transition" by Manton and Klinkhammer \cite{Manton}\cite{Arnold}, which is applicable to the baryogenesis as well as the leptogenesis \cite{Yanagida}, functioning at the electroweak (EW) scale.  The original scenario by Yoshimura \cite{Yoshimura} of generating baryon (B) number at the grand unified (GUT) scale, or the other scenarios generating B-number by the oscillation of a scalar field with B or L (lepton) number, such as the supersymmetric Affleck-Dine mechanism \cite{A-D}, the Dimopoulos-Susskind  \cite{D-S}, the Cohen-Kaplan \cite{K-K} and others \cite{ours 1}\cite{ours 2}\cite{ours 3}, still remain as plausible candidates for the baryogenesis scenario.

There are different ways to satisfy the two Sakharov's criterions, the existence of baryon number violating interaction and that of CP violation.  One way is to break explicitly B-conservation and CP-symmetry, and the other way is to break them by ``chiral anomaly" via a certain gauge field configurations with non-vanishing topological Chern number ($n_{C}$) or Chern-Simmons number ($n_{CS}$).  The sphaleron transition is an example of this anomaly mediated way, but other scenarios do exist.  

Indeed a while ago before the sphaleron was found, the author proposed another scenario   \cite{Sugamoto 1}, in which the neutrino mass (including L- and B-number violation processes) is induced by the monopolium configuration (monopole-anti-monopole dumbell system in SO(10) GUTs).  It is a generalization of Rubakov-Callan effect of monopole induced B-violation \cite{Rubakov}, but the configuration used was the monopolium found by Nambu \cite{Nambu}.  

In the standard model with a doublet Higgs field, the monopolium is the more natural solution than the isolated monopole obtained in the gauge model with a triplet Higgs field.  Nambu's monopolium (called monopole-anti-monopole dumb-bell system) obtained in the standard model, consists of a pair of monopole and anti-monopole, having a thin flux tube made of massive gauge boson $Z$, with solenoidal magnetic fields spreading out from the monopole and entering in the anti-monopole.  This thin flux tube is called $Z$-string in this paper.

The author's previous paper \cite{Sugamoto 1} was, however, so primitive regarding the quantitative analysis of the fermion number violation processes.  One crucial problem not solved completely then, was how to generate time-dependent electric field $\bm{E}(t, \bm{x})$, being parallel to the static magnetic field $\bm{B}(\bm{x})$ of the monopolium.  The generation of this electric fields is crucial so that it may have the non-zero Chern number: 
\begin{eqnarray}
\text{Chern number} = \frac{g^2}{8 \pi^2} \int d^4x \; \bm{E}(x) \cdot \bm{B}(x),
\end{eqnarray}
since the conservation of the L-handed fermion number current $J_L^{\mu}(x)=\bar{\psi}_L(x)\gamma^{\mu} \psi_L(x)$ is violated by the chiral anomaly:
\begin{eqnarray}
\partial_{\mu} J_L^{\mu}(x)= -\frac{g^2}{8 \pi^2}\bm{E}(x) \cdot \bm{B}(x).
\end{eqnarray}
The problem mentioned looks a trivial one, but the dynamics to generate electric field parallel to the magnetic field is not so easy to understood, that was exactly the question raised by Charlie Goebel to the author in 1982.

When we apply the chiral anomaly and the associated chirality flip phenomena for the chiral fermions in material science, we have no such difficulty.  The spin structure is effectively generated by the band structure of a certain material \cite{N-N}, the magnetic field and the electric field can be applied from outside in the laboratory.  Therefore, we can controll the positive and negative sign of the inner product, $(\bm{E} \cdot \bm{B})$, as we want.  It is also true for the heavy ion collision experiments in the laboratory.  However, the problem is serious in high energy physics and cosmology, where the sign of the inner product, $(\bm{E} \cdot \bm{B})$, can not be controlled by us.  

To overcome this difficulty, we take in this paper, not the monopolium but the dyonium, since the latter has electric field from the beginning supplied by the electric charges located at both end points of the system.  Then, it becomes possible to estimate the rate of L- and B-number violations as well as the neutrino mass.  Furthermore, we can discuss the physical implications, such as the number density and the interaction with other particles of this dyonium, in the early history of the universe.

Four decades ago, the author tried to use an analogy between the monopolium and the magnetosphere of the earth.  It is well-known the charged particles such as protons are trapped by the Van Allens bands \cite{Van Allen} \cite{Fermi-Rossi}.  The probability distribution of the trapped charged particle resembles the fermionic zero mode in the monopolium background.  As we know, however, the aurora (a kind of Rubakiv-Callan effect) happens equally at both poles, north and south, which shows the electric field generated by the trapped charged particle does not give a definite direction to itself, parallel or anti-parallel to the magnetic dipole field of the earth.  Thus, the study based on the analogy with the magnetosphere of the earth was not successful. This intuitive view may not be completely wrong, since restricting to the L-handed fermions, parallel direction of the spin to the magnetic field is favored, so that the direction of motion of the fermion is favored to be anti-parallel to the magnetic field.

Anyway, a long period has passed before arriving at a simple answer with dyonium to the Charlie Goebel's question. 

The paper is organized as follows:

In the introduction, the motivation to use dyonium instead of monopolium in the fermion number violation processes is described.  

In Section 2, the dyonium solution is given explicitly, in which electric field is excited parallel to the magnetic field, and the contribution of Z-string is included.

The standard method to estimate the fermion number violation processes is summarized in Section 3.  

In Section 4, a way to estimate the fermion zero modes is developed, in which the Dirac equation of $SU(2)_L \times U(1)_Y$ gauge theory can be reduced to a renormalization group-like equation for a single component function without isospin and spin.  

The last section is devoted to conclusion of this paper and discussion on the unsolved issues in it.

\section{Dyonium as a generalization of Nambu's monopolium}

Nambu gave a monopolium solution in the standard model (SM) ($SU(2)_L \times U(1)_Y$ gauge theory) with a doublet Higgs field \cite{Nambu}.  In his solution, the gauge fields have the arbitrariness represented by a function $a_{\mu}(x)$.  Using this arbitrariness we can obtain the dyonium solution, by adding the electric charges on both ends of the monopolium.

Here, we choose the gauge group $SU(2)_L \times U(1)_Y$, but it is not necessary to the SM.  It can be a subgroup embedded in the larger grand unified group such as $SO(10)$ and others, and prepare $N_d$ L-handed doublet fermions and $N_s$ singlet fermions.  The vacuum expectation value $v$ of the doublet Higgs 
\begin{eqnarray}
\langle \phi(x) \rangle = \begin{pmatrix} v \\ 0 \end{pmatrix}.
\end{eqnarray}
can be any value, not restricted to the SM value of $246/\sqrt{2}$ [GeV].

In this setup, we can study neutrino mass, as well as L- and B-number violation processes, depending on the choice of gauge group and fermion doublets. 

The simplest example is the SM with gauge group $SU(2)_L \times U(1)_Y$.  We have four L-handed fermion doublets for a generation, 
\begin{eqnarray}
\{\psi^{(0)}; \psi^{(1)}, \psi^{(2)}, \psi^{(3)}\}_{SM} =\left\{ \begin{pmatrix} \nu_e \\ e \end{pmatrix}_L;  \begin{pmatrix} u_1 \\ d_1 \;  \end{pmatrix}_L, \begin{pmatrix} u_2 \\ d_2 \;  \end{pmatrix}_L, \begin{pmatrix} u_3 \\ d_3 \;  \end{pmatrix}_L \right\}_{SM},  \label{eq0}
\end{eqnarray}
where indices $\{1, 2, 3\}$ represent color quantum numbers.

Our theory can be applicable to the models beyond the SM.  Some examples of this kind can be found in \cite{Sugamoto 1}, where the $SU(2)$ subgroup embedded in $SO(10)$ group is labeled by $(i, j)~(i, j=1-5)$, and the fermion doubles are the following ones:
\begin{eqnarray}
\{\psi^{(0)}; \psi^{(1)}, \psi^{(2)}, \psi^{(3)}\}_{ij} =\left\{ \begin{pmatrix} N_R^C \\ \psi_L(10)_{ij} \end{pmatrix};  \begin{pmatrix} \psi_L(10)_{kl} \\ \psi_R(5)^C_{m} \;  \end{pmatrix} \right\}_{i j},  \label{eq4}
\end{eqnarray}
where $(i,j,k,l,m)$ is an even permutation of $(1,2,3,4,5)$ and the $SU(2)_{(ij)}$ group is a subgroup of SO(10).\footnote{Three generators of the $SU(2)_{(ij)}$ group can be written by the creation and annihilation operators of the spinor representation of $SU(5)$, such that
$
\tau^1_{(ij)}= b_i^{\dagger}b_j^{\dagger}+b_jb_i, ~
\tau^2_{(ij)}= i(-b_i^{\dagger}b_j^{\dagger}+b_jb_i), ~
\tau^3_{(ij)}= 1- (b_i^{\dagger}b_i + b_j^{\dagger}b_j).
$
The $SU(5)$ multiplets $\psi_R(5)^C, \psi_L(10)$ consist of
$
\psi_R(5)^C=(\bar{d_i}, e, \nu )_L, ~\psi_L(10)=(\bar{u_i}, u_i, d_i, \bar{e})_L, ~(i=1, 2, 3; \text{color}).
$  Charge conjugation operation is defined by $\psi^{C}=i \gamma_2 \psi^*$, which gives $(\psi_L)^C= i \sigma_2 \psi_L$ for the L-handed fermion. }

Now, we can start with the following SM-like Lagrangian,  
\begin{eqnarray}
&&\mathcal{L}= -\frac{1}{4}\sum_{a=1}^3 (F^a_{\mu\nu})^2 -\frac{1}{4} (B_{\mu\nu})^2 + v^2 |D_{\mu}(A, B) \phi |^2 -\lambda v^4 (\phi^{\dagger} \phi-1)^2  \nonumber \\
&&+ \sum_{a=0}^{N_d-1} \bar{\psi}^{(a)}(x)_L \gamma^{\mu}  D_{\mu}(A, B) \psi^{(a)}(x)_L + \sum_{b=1}^{N_s} \bar{\psi}_s^{(b)}(x)_L \gamma^{\mu}  D_{\mu}(B) \psi_s^{(b)}(x)_L,
\end{eqnarray}
where 
\begin{eqnarray}
D_{\mu}(A, B)= \partial_{\mu} -i  \frac{\tau^a}{2} g A^a_{\mu}(x)-i  \frac{Y}{2} g' B_{\mu}(x), ~~D_{\mu}(B)= \partial_{\mu} -i  \frac{Y}{2} g' B_{\mu}(x).
\end{eqnarray}
Here, we have used the normalized Higgs field $|\phi|=1$ by the vacuum expectation value $v$, and the fermions are all represented by the L-handed ones, by applying properly the charge conjugation operation $(C)$.  The numbers of fermion doublets and singlets are denoted by $N_d$ and $N_s$, respectively.

The monopolium solution is approximately given by Nambu as a solution of the ``London equation",
\begin{eqnarray}
D_{\mu}(A, B) \phi(x)=0,  ~~\phi^{\dagger}(x)\phi(x)=1,
\end{eqnarray}
where $Y=-1$ for the Higgs field.\footnote{Here the up and down components of the Higgs field are exchanged from the usual choice.}  

It is natural to choose the configuration of  doublet Higgs field $\phi(x)$ spinor-likely as
\begin{eqnarray}
\phi(x) = \begin{pmatrix} \cos\frac{1}{2} \Theta(x) \\ \sin \frac{1}{2} \Theta(x)\; e^{i \varphi(x)} 
\end{pmatrix},  \label{Higgs doublet}
\end{eqnarray}
where $\cos\Theta = \cos \theta_1 - \cos \theta_2 + 1$ defined by $\theta_1$ and $\theta_2)$.  They are the polar angles between the $z$ axis and the position vectors of $x$, seen from the monopole and anti-monopole positions, located at $z = d/2$ and $z = - d/2$, respectively.  The useful variables to study the monopolium are the curvilinear but orthogonal coordinate system, $(\rho, \Theta, \varphi)$.  Here, $\rho$ is a magnetic potential of the monopole-anti-monopole system, \footnote{In this paper $\rho$ is chosen with dimension 1/[length] so that it may give naturally the potential. This definition differs by $d$ from the dimensionless $\rho$ in \cite{Sugamoto 1}.}
\begin{eqnarray}
\rho = \frac{1}{l_1} - \frac{1}{l_2}, 
\end{eqnarray}
given by $l_1 = [(z-d/2)^2+ r^2]^{1/2}, \; l_2=[(z+d/2)^2+ r^2]^{1/2}$ with $r = (x^2 + y^2)^{1/2}$.  Then, 
\begin{eqnarray}
\cos \Theta(x)= \frac{z-\frac{d}{2}}{l_1} - \frac{z+\frac{d}{2}}{l_2} +1.
\end{eqnarray}

The configurations of gauge fields are obtained by solving the London equation.  To solve it, we prepare two equations:
\begin{eqnarray}
\phi^{\dagger} (x) D_{\mu}(A, B) \phi(x)=0, ~\text{and}~\phi^{\dagger} (x) \tau^a D_{\mu}(A, B) \phi(x)=0,
\end{eqnarray}
from which the configurations of the gauge fields can be found.  The result reads 
\begin{eqnarray}
&&g A^a_{\mu}(x)= -\epsilon^{abc} (\phi^{\dagger} \tau^b \phi)\partial_{\mu}  (\phi^{\dagger} \tau^c \phi) + (\phi^{\dagger} \tau^a \phi) \left\{ - i \xi  (\phi^{\dagger} \overleftrightarrow{\partial}_{\mu} \phi) + a_{\mu}(x) \right\}, \\
&&g' B_{\mu}(x) = i \eta  (\phi^{\dagger} \overleftrightarrow{\partial}_{\mu} \phi) + a_{\mu}(x), \label{gauge fields}
\end{eqnarray}
where $\overleftrightarrow{\partial}=\overrightarrow{\partial}-\overleftarrow{\partial}$, $(\xi, \; \eta)$ are constants to be fixed with $\xi + \eta=1$, and $a_{\mu}(x)$ gives an arbitrariness of the solution.

The origin of the arbitrariness for $(\xi, \eta)$ and $a_{\mu}(x)$ is that the gauge field determined by the London equation is, a sum of $SU(2)$ and $U(1)$ gauge fields forming the massive gauge field for $Z$.  Indeed, in the London equation, only the sum $\xi + \eta=1$ appears and the terms of $a_{\mu}(x)$ are summed up cancel with each other.  

Therefore, the separation of the sum into $SU(2)$ and $U(1)$ parts has the arbitrariness.

Nambu has chosen the arbitrary function to be zero as $a_{\mu}(x)=0$, and derived the monopolium solution.

On the other hand in this paper, we utilize this arbitrariness and introduce the electric charges at both end points of the monopolium, which produces the dipole electric field parallel to the magnetic field.  Then, we have a ``dyonium solution''.  For this purpose we adopt the following ansatz,
\begin{eqnarray}
a_i(x) = 0, ~~a_0(x)= Q\rho(x). \label{choice of a}
\end{eqnarray}

Furthermore, we consider the oscillation (expansion and shrinkage) of the length $d(t)$ of the dyonium, in  estimating the temporal change of the Chern number.  Thus, $l_1, l_2$ and $\rho$ become time-dependent through the temporal change of the length $d(t)$.

Now, a straightforward calculation shows that the field strengths are given by
\begin{eqnarray}
&&g F^a_{\mu\nu}= (\phi^{\dagger} \tau^a \phi) \{ (\xi-1) f_{\mu\nu} +a_{\mu\nu}\}, \\
&&g' B_{\mu\nu}=  -  \eta f_{\mu\nu}  +  a_{\mu\nu}, \label{field strengths} 
\end{eqnarray}
where 
\begin{eqnarray}
&&f_{\mu\nu}(x)=-2i (\partial_{\mu} \phi^{\dagger} \partial_{\nu} \phi-\partial_{\nu} \phi^{\dagger} \partial_{\mu} \phi), \\
&&a_{\mu\nu}(x)=\partial_{\mu} a_{\nu}-\partial_{\nu} a_{\mu}.
\end{eqnarray}
For our Higgs configuration (\ref{Higgs doublet}), we have
\begin{eqnarray}
f_{\mu\nu}(x)=-  \partial_{\mu} (\cos\Theta(x)) \partial_{\nu}\varphi(x) +\partial_{\nu} (\cos \Theta(x)) \partial_{\mu}\varphi(x).
\end{eqnarray}
The corresponding electric field and magnetic field are given by
\begin{eqnarray}
&&f_{0i}=     \frac{r}{2} \left(\frac{1}{(l_1)^3} + \frac{1}{(l_2)^3} \right)  \dot{d}(t) (\bm{1}_{\varphi})^i \equiv f(r, z) \dot{d}(t) (\bm{1}_{\varphi})^i,  \\
&&\frac{1}{2}\epsilon_{ijk}f_{ij}=  \partial_k \rho,
\end{eqnarray}
where $\bm{1}_{\varphi}$, in the Cartesian coordinate system $(-\sin \varphi, \cos \varphi, 0)$, is a unit vector circulating the z-axis in the increasing direction of $\varphi$. In the derivation of the above, we use 
\begin{eqnarray}
&&\partial_t (\cos \Theta)= -\dot{d}(t) \times \frac{r^2}{2} \left(\frac{1}{(l_1)^3} + \frac{1}{(l_2)^3} \right),  \label{time derivative}\\
&& \begin{cases}
\bm{\nabla} \rho=\left( -r \left( \frac{1}{l_1^3}-\frac{1}{l_2^3} \right), 0, - \left( \frac{z-d/2}{l_1^3}-\frac{z+ d/2}{l_2^3} \right) \right)_{\text{cylind}}, \\
\bm{\nabla}\cos \Theta=\left( - r \left( \frac{z-d/2}{l_1^3}-\frac{z+ d/2}{l_2^3} \right), 0, r^2 \left( \frac{1}{l_1^3}-\frac{1}{l_2^3} \right) \right)_{\text{cylind}}, \\
 \bm{\nabla} \varphi=\left(0, \frac{1}{r}, 0 \right)_{\text{cylind}},
 \end{cases} \label{space derivative}
 \end{eqnarray}
giving
\begin{eqnarray}
&& \bm{\nabla}\rho \cdot \bm{\nabla} \cos \Theta=0, \bm{\nabla}\cos \Theta \cdot \bm{\nabla} \varphi=0, \bm{\nabla} \varphi \cdot \bm{\nabla}\rho=0, \\
&& \bm{\nabla}\cos \Theta \times \bm{\nabla} \varphi=   \bm{\nabla} \rho, \;\bm{\nabla}\rho \times \bm{\nabla} \varphi= - \frac{1}{r^2} \bm{\nabla} \cos \Theta, \bm{\nabla}\rho \times \bm{\nabla}\cos \Theta=  r^2 (\bm{\nabla} \rho)^2 \bm{\nabla} \varphi. ~~~~~
\end{eqnarray}
These relations imply that the ``dipolar system'' such as dyonium (and monopolium), can be quite naturally described by a curved coordinate system, given by $(\rho, \varphi, \cos \Theta)_{\text{dipolar}}$ \cite{Sugamoto 1}.
 
Here, however, the cylindrical coordinate system $(dr, rd\varphi, dz)_{\text{cylind}}$ is used, (since it is more familiar than the dipolar system) for which the spacial derivative becomes, 
\begin{eqnarray}
\bm{\nabla} =\left(\frac{\partial}{\partial r}, \frac{1}{r} \frac{\partial}{\partial \varphi}, \frac{\partial}{\partial z} \right)_{\text{cylind}},
\end{eqnarray}
and also $\bm{1}_{\varphi}=(0, 1, 0)_{\text{cylind}}$.

For our choice of $a_{\mu}(x)$, we have 
\begin{eqnarray}
a_{0i}(x)=-Q\partial_i\rho, ~~a_{ij}(x)=0.
\end{eqnarray}

Thus the field strengths have simple expressions which helps to understand the structure of the dynonium, 
\begin{eqnarray}
&&g \bm{B}^a=(\phi^{\dagger}\tau^a \phi) (\xi-1)  \bm{\nabla}\rho, \label{eq30}\\
&&g \bm{E}^a= (\phi^{\dagger}\tau^a \phi)\left\{ (\xi-1)r f(r, z) \dot{d}(t) \bm{\nabla} \varphi -Q \bm{\nabla} \rho \right\}, \label{eq31}\\
&& g' \bm{B}'=-\eta  \bm{\nabla} \rho, \label{eq32}\\
&& g' \bm{E}'=(-\eta)  r f(r, z) \dot{d}(t) \bm{\nabla} \varphi - Q \bm{\nabla} \rho, \label{eq33}
\end{eqnarray}
where 
\begin{eqnarray}
f(r, z)=\frac{r}{2} \left( \frac{1}{l_1^3}-\frac{1}{l_2^3} \right),
\end{eqnarray}
and the magnetic field and electric field for the $U(1)_Y$ part are denoted with prime.

\subsection{Gauge potentials} 
On the other hand, to express the gauge potentials explicitly, we have to introduce three orthonormal vectors in the $SU(2)$ iso-space \cite{Sugamoto 1}. 
A typical iso-vector is the triplet Higgs field $(\phi^{\dagger} \tau^a \phi)$ which is a orthonormal basis, denoted by $n^a_{1}$,
\begin{eqnarray}
n^a_{1}  \equiv (\phi^{\dagger} \tau^a \phi) = (\sin \Theta \cos \varphi, \; \sin \Theta \sin \varphi, \; \cos \Theta)_{a=1-3},
\end{eqnarray}
where $a=1, 2, 3$ denotes the direction of iso-vector in the iso-space.

The remaining two unit vectors are, respectively, 
\begin{eqnarray}
&&n^a_{2}=(\cos \Theta \cos \varphi, \; \cos \Theta \sin \varphi, \; -\sin \Theta)_{a=1-3}, \\
&&n^a_{3}=(-\sin \varphi, \; \cos \varphi, \; 0)_{a=1-3}.
\end{eqnarray}
They satisfy the orthonormality condition as iso-vectors,
\begin{eqnarray}
\epsilon_{abc} (n_j)^b (n_k)^c= \epsilon_{ijk} (n_i)^a,
\end{eqnarray}
and the derivative of one of them can be expanded in the remaining two,
\begin{eqnarray}
\partial_{\mu}(n_1)^a=(n_2)^a \partial_{\mu} \Theta + (n_3)^a \partial_{\mu} \varphi.
\end{eqnarray}

Thus, we have
\begin{eqnarray}
&& (\phi^{\dagger} \tau^a \phi) \equiv n_{1}^a, ~ \partial_{\mu}(\phi^{\dagger} \tau^a \phi)=n_{2}^a \partial_{\mu} \Theta +n_{3}^a \sin \Theta \; \partial_{\mu} \varphi,  ~\text{and} \\
&& (\phi^{\dagger} \overleftrightarrow{\partial_{\mu}}\phi)=-i  (1-\cos \Theta) \partial_{\mu}\varphi.
\end{eqnarray}

Then, the gauge potentials become
\begin{eqnarray}
&&g A^a_{\mu}(x)=n^a_{1}\left\{ -\xi (1- \cos \Theta)\partial_{\mu} \varphi + a_{\mu}(x) \right\}  +n^a_{2} \sin \Theta \partial_{\mu} \varphi + n^a_{3} \frac{1}{\sin \Theta} \partial_{\mu} (\cos \Theta), \\
&& g' B_{\mu}(x) = \eta (1- \cos \Theta)\partial_{\mu} \varphi +  a_{\mu}(x).
\end{eqnarray}
Using (\ref{time derivative}), (\ref{space derivative}) and (\ref{choice of a}), we have obtained the more explicit expressions for the gauge potentials,
\begin{eqnarray}
&&g A^a_{0}(x)=n^a_{1} Q \rho(x) + n^a_{3} \frac{1}{\sin \Theta} f(r, z) \dot{d}(t), \label{eq44} \\
&&g \bm{A}^a(x)= -  \{ \left( n^a_{1} (-\xi) (1- \cos \Theta)\bm{\nabla}\varphi + n^a_{2} \sin \Theta \right)  \bm{\nabla}\varphi  + n^a_{3} \frac{1}{\sin \Theta} (\bm{\nabla} \cos \Theta)\}, \label{eq45} \\
&& g' B_0(x)=Q\rho(x), ~~g' \bm{B}(x)=-  \eta (1- \cos \Theta) (\bm{\nabla}\varphi). \label{eq46}
\end{eqnarray}

About the components of the gauge fields, see the following Subsection 2.2.

\subsection{Determination of parameters $(\xi, \eta)$ with $\xi+\eta=1$}

Using the termnology in SM, the neutral gauge fields in our model consist of a massive ``Z-boson" ($Z$) and a massless ``photon'' ($\gamma$).  These components are, respectively,  
\begin{eqnarray}
&&\sqrt{g^2 +g^{'2}} F^{Z}_{\mu\nu}=-(\phi^{\dagger} \tau^a \phi) g F^a_{\mu\nu}+ g' B_{\mu\nu}=0, ~\text{and} \\
&& \sqrt{g^2 +g^{'2}} F^{\gamma}_{\mu\nu}= + (\phi^{\dagger} \tau^a \phi) g' F^a_{\mu\nu}+ g B_{\mu\nu}= \left(\frac{g^2+g^{'2}}{gg'} \right) \left\{ -\eta f_{\mu\nu}(x) + a_{\mu\nu}(x) ) \right\} \ne 0,~~~~~
\end{eqnarray}
where the zero of the first equation means the massive $Z$ filed is confined inside a thin flux tube connecting a dyon and an anti-dyon, while the non-zero of the second equation means the massless photon field spreads out to form the magnetic and the electric dipole fields. 

The configurations of gauge potentials are given in (42)--(44).

The singularities for the gauge potential appear generally inside the tube connecting monopole and anti-monopole at $r=0, \; \Theta=\pi$.  

For the $Z$ flux, in the approximation using the London equation, we have  
\begin{eqnarray}
\sqrt{g^2+g^{'2}} \bm{A}^{(Z)}= - g(\phi^{\dagger}\tau^a \phi) \bm{A}^a+g' \bm{B} \approx  \underbrace{- i (\phi^{\dagger}\overleftrightarrow{\nabla} \phi)}_{\text{by London eq.}} + \underbrace{0}_{\text{correction}}=  \frac{1-\cos \Theta}{r} \bm{1}_{\varphi} + 0,  \label{by London eq.}
\end{eqnarray}
so that the gauge field for $Z$ looks to have a singularity for $r \to 0$ in this approximation.  However, this is not correct when we take into account the decrease of the Higgs field in the flux tube.  This is well-known from the Meissner effect in superconductivity and gauge models, in which the singularity of the gauge potential disappears due to the rapid decrease of the Higgs expectation value there.   Nambu stated this situation as the singularity disappears due to the smearing by the Higgs field.

As for the gauge potential for $\gamma$, the singularity can not be smeared out by the Higgs field, so that the singularity remains at $r=0, \; \Theta=\pi$,
\begin{eqnarray}
\sqrt{g^2+g^{'2}} \bm{A}^{(\gamma)}= + g'(\phi^{\dagger}\tau^a \phi) \bm{A}^a+g \bm{B} = \left\{ -\xi \frac{g'}{g} + \eta \frac{g}{g'} \right\} \frac{1-\cos \Theta}{r} \bm{1}_{\varphi}.
\end{eqnarray}
Therefore, we have to impose the following condition to discard the singularity, 
\begin{eqnarray}
-\frac{g'}{g} \xi + \frac{g}{g'} \eta=0,
\end{eqnarray}
which fixes the parameters $(\xi, \eta)$ as follows:
\begin{eqnarray}
\xi= \frac{g^2}{g^2+g^{'2}}=\cos^2 \theta_W, ~~\eta= \frac{g^{'2}}{g^2+g^{'2}}=\sin^2 \theta_W.
\end{eqnarray}
Here, we have used the notation $\theta_W$, but it is the mixing angle between massive and massless gauge bosons in the dyonium model, and is not necessary to take the SM value.

On the other hand, the time component of the gauge potential for photon is
\begin{eqnarray}
\sqrt{g^2+g^{'2}} A_0^{\gamma}=  + g'(\phi^{\dagger}\tau^a \phi) A_0^a+g B_0 = \frac{g'}{g} Q \rho(x),
\end{eqnarray}
which is singular at the positions of dyon ($l_1=0$) and anti-dyon ($l_2=0$), but these singularities are normal ones for the point charges.  If necessary, they can be smeared out by introducing a finite size $\delta$ to dyon and anti-dyon.

Physical picture of the monopolium stated by \cite{Nambu} is the following: A monopole and an anti-monopole are connected by a thin tube having a width $(gv)^{-1}$ of the symmetry breaking scale.  The $SU(2)$ flux of $4\pi/g$ goes out (in) from the monopole (to the anti-monopole), the part of which $\eta(4\pi/g)$ is radiated as a dipole field, while the remaining part of which $\xi (4\pi/g)$ flows inside the tube.  On the other hand, the $U(1)$ flux of $\eta(4\pi/g')$ forms a solenoid without end-points, which gives both a dipole field outside and a flux inside the tube. In our case of dyonium, we have additional electric chares $Q'_e=Q/g'$ at the monopole position and $-Q'_e$ at the anti-monopole position.  From the electric chafes, the dipole electric field is radiated.
\subsection{Modification of London equation near Z-string}
We have to know the features of magnetic flux and gauge potential near the $Z$-string more precisely.  For this purpose, we modify the London equation near the Z-string located on the $z$-axis from $-d/2$ to $d/2$.  The Z-flux flows inside the thin tube along it.  

The equations of motion of our $SU(2)_L \times U(1)_Y$ model are, precisely, 
\begin{eqnarray}
&&\circ \; \left(D_{\mu}(A, B)\right)^2 \phi(x) = -\lambda v^2 (\phi^{\dagger}\phi-1)\phi(x), \label{EOM 1} \\
&&\circ \; \left( \delta^{ac} \partial_{\mu} - g \epsilon^{abc} A^b_{\mu} \right) F^{c, \mu\nu}(x) = - i g v^2 \left( \phi^{\dagger} \frac{\tau^a}{2} D_{\mu}(A, B) \phi - (D_{\mu}(A, B)  \phi)^{\dagger} \frac{\tau^a}{2} \phi  \right), \label{EOM 2} \\
&&\circ ~~ \partial_{\mu}  B^{\mu\nu}(x) = - i g' v^2 \left( \phi^{\dagger}  \frac{-1}{2}  D_{\mu}(A, B) \phi -   \frac{-1}{2} (D_{\mu}(A, B) \phi)^{\dagger} \phi \right).  \label{EOM 3}
\end{eqnarray}

London equation is an approximation, in which the  Higgs potential in the right hand side of (\ref{EOM 1}) vanishes.  This is valid when the expectation value of the Higgs doublet $\langle \phi(x)^{\dagger} \phi(x)\rangle$ takes unity everywhere, when normalized by a constant $v^2$.  This is, however, not valid near the $Z$-string.
Accordingly, London equation is to be modified near the $Z$ string, which is well-known from Meissner effect in superconductivity by Ginzburg and Landau \cite{G-L}, and the string-like solution in gauge theories by Nielsen and Olesen \cite{N-O}.

Therefore, we modify the Higgs field $\phi(x)$ and the gauge field $\bm{A}^{(Z)}(x)$ of $Z$, from the previous ones based on the London equation, to the following ones:
\begin{eqnarray}
&&\circ~\phi(x)= H(x) \phi_0(x) = \left(\underbrace{1}_{\text{by London eq.}} + \underbrace{\bar{H}(x)}_{\text{correction}} \right) \phi_0(x), ~~\text{and}\\
&&\circ~\sqrt{g^2+g^{'2}}\bm{A}^{(Z)}(x) = -(\phi^{\dagger} \tau^a  \phi) g \bm{A}^a + (\phi^{\dagger} \phi) g'\bm{B} \\
&&=  \underbrace{\frac{-i \left(\phi(x)^{\dagger} \overleftrightarrow{\nabla} \phi(x)\right)}{(\phi^{\dagger} \phi)}}_{\text{ by London eq.}}  + \underbrace{\sqrt{g^2+g^{'2}} \bar{\bm{A}}(x)}_{\text{correction}},
\end{eqnarray}
where $\phi_0(x)$ is the normalized Higgs configuration used in Eq.(10).  The last equation is the modification of the previous Eq.(\ref{by London eq.}). 


Then, the equations of motion (\ref{EOM 1}), (\ref{EOM 2}), (\ref{EOM 3}) can be written only in terms of the corrections $\bar{H}$ and $\bar{\bm{A}}$, with the definition of $M_Z=\sqrt{g^2 +g^{'2}} \; v/\sqrt{2}$ and $m_H= 2\sqrt{\lambda}v$, namely, 
\begin{eqnarray}
&&\circ \; \bm{\nabla}^2 \bar{H}(x) =  \frac{1}{4}  (g^2+g^{'2}) \bar{\bm{A}}(x)^2 \left(1+ \bar{H}(x)\right) + m_H^2 \left( \bar{H} (x) + \frac{3}{2} \bar{H}(x)^2+ \frac{1}{2} \bar{H}(x)^3 \right),  \\
&& \circ \; \nabla \times (\nabla \times \bar{\bm{A}}(x))= m_Z^2 \left(1+ \bar{H}(x)\right)^2 \bar{\bm{A}}(x),
\end{eqnarray}
where $\frac{-i (\phi^{\dagger} \overleftrightarrow{\nabla} \phi)}{(\phi^{\dagger} \phi)}= \frac{1- \cos \Theta}{r} \bm{1}_{\varphi}$ and the identity $(\tau^a n^a_1) \phi(x)=\phi(x)$ have been used.  The former relation confirms $\bm{A}^{(Z)}$ in Eq.(49).

In solving these equations, it is not so easy to use the coordinate system $(\rho, \cos \Theta)$, even on a plane with a fixed $\varphi$.\footnote{ The Laplacian to use is, $
\nabla_{(2D)}^2= r(\nabla \rho)^2 \left\{ \partial_{\rho} \frac{1}{r} \partial_{\rho} +  \partial_{\cos \Theta} r \partial_{\cos \Theta} \right\} $, where $r$ should be expressed in terms of $(\rho, \cos \Theta)$.  This is at least possible numerically.} (See Discussion on the issue 2) about the usage of the rotating elliptic coordinate system.)   

Therefore, what we can do here is to assume the infinitely long dyonium, by considering the dependency only on $r$ and ignoring $z$.  Then, our equations become identical to those studied by Nielsen and Olesen in gauge theories.

The equations of motion become\footnote{In deriving the second equation, Stokes theorem is applied for a circle $S$ and its boundary curve $C$, surrounding the $Z$-string with radius $r$,
$$ \int\int_{S(r)}  (\nabla \times \bm{A}) \cdot d\bm{S} = \oint_{\partial S=C(r)}  \bm{A} \cdot d\bm{x}, ~~\text{or}~~ (\nabla \times \bm{A})_z= \frac{1}{r} \partial_r (r A(r)).$$}
\begin{eqnarray}
&& \left(\partial_r^2 + \frac{1}{r} \partial_r \right) \bar{H}(r) =  \frac{1}{4}  (g^2+g^{'2}) \bar{\bm{A}}(r)^2 \left(1+ \bar{H}(r)\right) + m_H^2 \left( \bar{H} + \frac{3}{2} \bar{H}^2+ \frac{1}{2} \bar{H}^3 \right),   \\
&&  \partial_r \left( \frac{1}{r} \partial_r \left( r \bar{A}(r) \right) \right)= \left( \partial_r^2 + \frac{1}{r} \partial_r  -  \frac{1}{r^2} \right)  \bar{A} (r) =m_Z^2 \left(1+ \bar{H}(r)\right)^2 \bar{A}(r).
\end{eqnarray}

In the linear approximation, the solution is well-known since \cite{N-O} as follows:
\begin{eqnarray}
\bar{A}(r) = a \;  m_Z  K_1(m_Z r), ~\text{and}~ \bar{H}(r) = b \; K_0(m_H r), \label{solution}
\end{eqnarray}
where $K_m(r)$ is the modified Bessel function, satisfying
\begin{eqnarray}
\left( \partial_r^2 + \frac{1}{r} \partial_r  -  \frac{m^2}{r^2} \right)  K_m(r) = K_m(r),
\end{eqnarray}
and the asymptotic behavior for $r \to \infty$ is, in general,
\begin{eqnarray}
K_m(r) \sim \sqrt{\frac{\pi}{2r}}e^{-r} \left\{ 1+ \frac{m^2-1/4}{2r} + O\left(\frac{1}{r^2}\right) \right\} ~(r \to \infty). 
\end{eqnarray}
As for the behavior of $r \to 0$, $K_1(r)$ is simple, but $K_0(r)$ is a little delicate,
\begin{eqnarray}
K_1(r \to 0) \sim  \frac{1}{r}, ~~\text{but} ~~ K_0(r \to 0) \sim  -b(\gamma_E + \ln (r/2)),
\end{eqnarray}
where $\gamma_E$ is the Euler's constant.

We will fix $a$ in the next subsection, but leave $b$ free.  
 
Now, the gauge field coming from the $Z$-string can be expressed by
\begin{eqnarray}
&& \bm{A}^{a(Z)} (x) = (n^a_1 \bm{1}_{\varphi})  \left\{  \frac{1- \cos \Theta}{r \sqrt{g^2+g^{'2}}} + a \; m_Z  K_1(m_Z r) \right\}.  \label{gauge field for Z}
\end{eqnarray}

We need to know the tension of $Z$-string, the energy stored per unit length of the string, which was estimated in \cite{N-O}: 
\begin{eqnarray}
E_{\text{Z-string}}= K d, ~\text{with} ~~  K  = c \; m_Z^2. \label{string tension}
\end{eqnarray}
 In the next subsection, we will estimate the parameter $c$.

\subsection{Estimation of Z-string parameters, $a$ and $c$}

To estimate the parameters $a$, we use the property that the gauge field $\bm{A}^{(Z)}$ is smeared so that it is free from the singularity at $r=0$.  Then, we have
\begin{eqnarray}
A^{(Z)} ~\underset{r \to 0}{\longrightarrow} ~\frac{2}{r \sqrt{g^2+g^{'2}}} + a \frac{1}{r} =0, ~~\text{giving}~~a= \frac{-2}{ \sqrt{g^2 +g^{'2}}}.
\end{eqnarray}

Now, we have finally obtained the expression for the gauge field of $Z$-string, as follows:
\begin{eqnarray}
\bm{A}^{(Z)a}(x)= n_1^a A^{(Z)}(x)\bm{1}_{\varphi}, ~~\text{and}~~ A^{(Z)} (x)= \frac{1}{\sqrt{g^2+g^{'2}}}\left( \frac{1-\cos \Theta }{r} - 2m_Z \; K_1(m_Z r)\right). \label{gauge field for Z}
\end{eqnarray}

To fix $c$, the coefficient of Z-string tension, we have to estimate the energy per unit length, coming from the magnetic field $\nabla \times \bm{A}^{(Z)}$,
\begin{eqnarray}
(\nabla \times \bm{A}^{(Z)})_z=   \frac{2m_Z^2}{\sqrt{g^2+g^{'2}}} \;  K_0 (m_Z r).
\end{eqnarray}
Thus, the tension $K= E_{\text{Z-string}}(d)/d =c m_Z^2$, or the parameter $c$, becomes
\begin{eqnarray}
c = \frac{4 \pi }{ (g^2 +g^{'2})}\int_0^{\infty} r dr \; K_0(r)^2=\frac{2 \pi }{ (g^2 +g^{'2})},
\end{eqnarray}
where $c$ is finite, even if $K_0(r)$ diverges logarithmically at $r \to 0$.\footnote{~Shiro Komata informed the author of the estimation of $c$ in their private communication:
$$\int_0^{\infty} r \; dr  K_0(r)^2 = \frac{1}{2} \left[ r^2 \left\{ K_0(r)^2 - K_1(r)^2 \right\} \right]_0^{\infty}= \frac{1}{2}.$$} 

In this way, the parameters $a$ and $c$ for $Z$-string have been determined.   

\subsection{Charge quantization condition for dyon and fixing of $Q$}
We have to impose the quantization condition between electric and magnetic charges, following Schwinger \cite{Schwinger}.

We know that no magnetic charge exists for $U(1)$ part (in QED), so that we consider both magnetic charge $Q^a_m$ and electric charge $Q^a_e$ for $SU(2)$ part, while only the electric charge $Q'_e$ for $U(1)$ part, giving a list of charges $(Q^a_m, Q^a_e; Q'_e)$.

In our dyonium we have the following field strengths for $SU(2)$ and $U(1) $ parts, in the static situation without time-dependency: 
\begin{eqnarray}
&&g \bm{B}^a=(\phi^{\dagger}\tau^a \phi) (-\eta) ( +  \bm{\nabla}\rho), ~g \bm{E}^a= (\phi^{\dagger}\tau^a \phi) Q (-\bm{\nabla} \rho ), \\
&& g' \bm{B}'=-\eta ( +  \bm{\nabla} \rho), ~ g' \bm{E}'=  Q (-\bm{\nabla} \rho).
\end{eqnarray}

This shows the solenoid-like dipole electric and magnetic fields, spreading out from the dyonium.  The magnetic flux inside the thin tube coming from the $SU(2)$ monopole is ignored in this expression.  This missing part can be recovered by taking a pure $SU(2)$ model with $\xi=1$, 
\begin{eqnarray}
(g \bm{B}^a)_{SU(2)\text{monopole}}= (\phi^{\dagger}\tau^a \phi) (-1) ( + \bm{\nabla}\rho).
\end{eqnarray}

Now, we have
\begin{eqnarray}
\left(\bm{B}^a_{SU(2)\text{monopole}}, \bm{E}^a; \bm{E}' \right)=\left(\frac{ n^a_{1}}{g}, \frac{n^a_{1}Q}{g}, \frac{Q}{g'} \right) (-\bm{\nabla} \rho),
\end{eqnarray}
where $n^a_{1}=(\phi^{\dagger}\tau^a \phi)$.

Using $\bm{\nabla}(-\bm{\nabla}\rho)=4\pi \left\{\delta^{(3)}(l_1)-\delta^{(3)}(l_2)\right\}$, we can read magnetic and electric charges $(Q_m, Q_e)$ from $\bm{\nabla} \bm{B}=4\pi Q_m \delta^{(3)}(\bm{x}-\bm{x}_m)$ and $\bm{\nabla} \bm{E}=4\pi Q_e \delta^{(3)}(\bm{x}-\bm{x}_e)$. The charges at $z=\pm \frac{d}{2}$ are
\begin{eqnarray}
(Q^a_m, Q^a_e; Q'_e)= \pm \left( \frac{n^a_{1}}{g}, \frac{n^a_{1}Q}{g}, \frac{Q}{g'} \right).
\end{eqnarray}

It is known that the electric charge and the magnetic charge $(Q_e, Q_m)$ of the dyon should be quantized \cite{Schwinger}, in order to define the non-singular angular momentum between two dyons with $(Q_e, Q_m)_1$ and $(Q_e, Q_m)_2$.  The condition is
\begin{eqnarray}
(Q_{e})_1 (Q_{m})_2 - (Q_{e})_2 (Q_{m})_1 = (\hbar c) \times (\text{integer}),
\end{eqnarray}
where $(\hbar c)$ is included to clarify the dimension of electric and magnetic charges.\footnote{~ Dimensions are $[e]=[g]=[g']=[\hbar c]^{1/2}$, since in QED $\alpha_e=\frac{e^2}{4\pi \hbar c}=(137.0359 \cdots)^{-1}$ is dimensionless.}

In applying this condition to $SU(2)$ monopole and charge, if there exists single type of dyoniums, then $Q$ is arbitrary.  However, if we consider two kinds of dyoniums exist; in one kind of dyonium, dyon has a charge $(Q^a_m, Q^a_e; Q'_e)_1=\left(\frac{n^a_{1}}{g}, \frac{n^a_{1}Q}{g}, \frac{Q}{g'} \right)$, while in the other kind, dyon has a charge $(Q^a_m, Q^a_e; Q'_e)_2=\left(\frac{n^a_{1}}{g}, -\frac{n^a_{1}Q}{g}, -\frac{Q}{g'} \right)$; then the quantization condition determines $Q$ as
\begin{eqnarray}
Q= \frac{g^2}{2}  \times (\text{integer}).
\end{eqnarray} 

Then, we can choose as an example, $Q=\frac{g^2}{2}$, which determines 
\begin{eqnarray}
(Q^a_m, Q^a_e; Q'_e)= \pm \left( n^a_{1}\frac{1}{g}, \pm n^a_{1}\frac{g}{2}, \frac{g}{2 \tan \theta_W} \right).
\end{eqnarray}

Now, the dyonium configuration in the $SU(2)_L \times U(1)_Y$ gauge model has been established.  

\section{A method to estimate L- and B-number violation processes}

In the previous section, we have succeeded in exciting the time-dependent electric fields of $SU(2)$ part $\bm{E}^a(x, t)$ and $U(1)$ part $\bm{E}'(x, t)$, so that they can be parallel to the magnetic fields $\bm{B}^a(x, t)$ and $\bm{B}'(x, t)$.  The important point here is the electric field and magnetic field are always parallel or anti-parallel, having the definite sign (positive or negative) of $\bm{E}(x, t) \cdot \bm{B}(x, t)$.  Then, Chern number and fermion number are ``monotonously increasing or decreasing''. This is the reason why dyonium was chosen in this paper.  Even if the generation of electric field is expected for a given configuration of magnetic fields, it is not easy to control the direction of the electric field, parallel or anti-parallel to the magnetic field.  Stochastically at finite temperature, the direction of electric field is arbitrary at each time, resulting no net L- and B- number generation, unless the external bias is introduced by the non-vanishing chemical potential, $(\mu_B$, $\mu_L$ or else)$\ne 0$.

\subsection{Chiral anomaly and Chern number}
In the presence of parallel or anti-parallel electric fields, the ``chiral anomaly" is induced radiatively.  

To make definite the statement in Introduction, we introduce the total fermion number current for the left-handed fermion doublets, $J_{(d)L}^{\mu}(x)= \sum_{a=0}^{N_d-1} \bar{\psi}_L^{(a)}(x)\gamma^{\mu} \psi_L^{(a)}(x)$, and fermion singlets $J_{(s)L}^{\mu}(x)= \sum_{a=0}^{N_s-1} \bar{\psi}_L^{(a)}(x)\gamma^{\mu} \psi_L^{(a)}(x)$, where the conservation of them are violated via the chiral anomaly for $SU(2)$ and $U(1)$ groups,  
\begin{eqnarray}
\partial_{\mu}J_{(d)L}^{\mu}(x)= -\frac{g^2N_d}{32 \pi^2} \left(F^a_{\mu\nu}(x) \tilde{F}^{a\mu\nu}(x)\right)=-\frac{g^2N_d }{8 \pi^2}\left( \bm{E}^a(x) \cdot \bm{B}^a(x) \right), \\
\partial_{\mu}J_{(s)L}^{\mu}(x)= -\frac{g'^2N_s}{32 \pi^2} \left(B_{\mu\nu}(x) \tilde{B}^{\mu\nu}(x)\right)=-\frac{g^{'2}N_s }{8 \pi^2}\left( \bm{E}'(x) \cdot \bm{B}'(x) \right).
\end{eqnarray}

The expression of gauge fields obtained in (\ref{eq30})--(\ref{eq33}) gives the non-vanishing Chern numbers for the dyonium, 
\begin{eqnarray}
&&q=\text{Chern}(SU(2)) \equiv \frac{g^2}{8 \pi^2} \int d^4 x \left( \bm{E}^a(x) \cdot \bm{B}^a(x) \right) 
=    \left(\frac{e}{4 \pi}\right)^2 \int d^4 x \; (-\bm{\nabla} \rho(x) )^2, ~~~\\
&&q'=\text{Chern}(U(1)) \equiv \frac{g'^2}{8 \pi^2} \int d^4 x \left( \bm{E}'(x) \cdot \bm{B}'(x) \right)  =    \left(\frac{e}{4 \pi}\right)^2 \int d^4 x \; (-\bm{\nabla} \rho(x) )^2,
\end{eqnarray}
where $e=g \sin \theta_W$ is the ``electric charge".
Therefore, the four dimensional integral of the anomaly equation yields
\begin{eqnarray}
\left[ n_F\right]_{t=-\infty}^{t=\infty} = \left[ \sum_{a=0}^{N_d-1} n^{(a)}_F\right]_{t=-\infty}^{t=\infty} +\left[ \sum_{b=0}^{N_s-1} n^{'(b)}_F\right]_{t=-\infty}^{t=\infty} =- (N_d  \; q + N_s \; q') \; (\ne 0), 
\end{eqnarray}
where $n^{(a)}_F$ and $n^{'(b)}_F$ stands for, respectively, the number of fermions for $a$-th doublet and $b$-th singlet, 
\begin{eqnarray}
n^{(a)}_F = \int d^3 x \;  \psi_L^{(a)}(x)^{\dagger} \psi_L^{(a)}(x), ~n^{'(b)}_F = \int d^3 x \;  \psi_L^{(b)}(x)^{\dagger} \psi_L^{(b)}(x).
\end{eqnarray}

\subsection{Dyonium action}
As for the action of dyonium $S_{\text{dyonium}}$, the contribution from the dipole fields is given in (\ref{eq30})--(\ref{eq33}), but the other contribution from the thin Z-string should also be included.  This additional term comes from the linear potential of the $Z$-string,
\begin{eqnarray}
S_{Z-\text{string}}= -\int d^4x \; V(x)_{Z-\text{string}}=-\int dt \; K d(t),
\end{eqnarray}
where the constant $K$ gives the tension of Z-string, given by (\ref{string tension}) in Section 2.3, 
\begin{eqnarray}
K= \frac{2\pi m_Z^2}{g^2+g'^2}.
\end{eqnarray}

Then, we have
\begin{eqnarray}
&&S_{\text{dyonium}}  \nonumber \\
&&=\frac{1}{e^2} \int d^4 x \left\{ \left(- \sin^4 \theta_W + \frac{g^4}{4}\right) (-\bm{\nabla}\rho)^2 +  \frac{g^4}{4} f(r, z)^2 \dot{d}(t)^2 \right\} -\int dt \; K d(t),
\end{eqnarray}
with $e=gg'/\sqrt{g^2 + g^{'2}}=g \sin \theta_W$.

\subsection{Temporal development of Chern number and action}
To estimate the temporal development of Chern number and action, we need to know:
\begin{eqnarray}
\int_D d^3 x (-\bm{\nabla}\rho(x))^2 = 2 \times  \left[4\pi l^2 \left(\frac{1}{l}-\frac{1}{d(t)} \right) \times \frac{1}{l^2}\right]_{l=\delta} =   8\pi  \left(\frac{1}{\delta}-\frac{1}{d(t)} \right).
\end{eqnarray}
Here, $D$ is the three dimensional domain, excluding the two small spheres with radius $\delta$ around dyon and anti-dyon.  The $\delta$ is the same order as $1/m_W$ or $1/m_Z$. 

Now we have
\begin{eqnarray}
q= q'= \int dt \left(\frac{1}{\delta}-\frac{1}{d(t)} \right).
\end{eqnarray} 

Here, we note again that Chern number is a time integral of a positive function, so that  ``Chern number is generated monotonously in time''. This is due to the dynonium, having definite electric and magnetic charges at both end points.

As for the action $S_A$ of the dyonium, the dipole type $\frac{1}{2}\left\{(\bm{B}^a)^2+(\bm{E}^a)^2+(\bm{B}')^2+(\bm{E}')^2\right\}$ energy and the string type energy $\frac{1}{2}(\bm{B}^{(Z)})^2$ contribute to it, 
\begin{eqnarray}
&&~~~S_{\text{dyonium}} \nonumber \\
&&= \int dt \left\{\frac{8\pi}{e^2} \left( - \sin^4 \theta_W+ \frac{g^4}{4} \right) \left(\frac{1}{\delta}-\frac{1}{d(t)} \right) +   \frac{g^4}{4e^2} \left(\int d^3x \; f(r, z)^2\right) \dot{d}(t)^2 - K d(t) \right\}. ~~~~~~~~ 
\end{eqnarray}
Introducing the constants, $M$ and $C$,
\begin{eqnarray}
&&M=\frac{g^4}{2e^2} \left(\int d^3x \; f(r, z)^2\right), ~\text{and} \\
&&C=\frac{8\pi}{e^2} \left( \sin^4 \theta_W - \frac{g^4}{4} \right),
\end{eqnarray}
we have arrived at
\begin{eqnarray}
&&S_{\text{dyonium}}= \int dt \left\{ \frac{M}{2}\dot{d}(t)^2 -V_{\text{dyonium}}(d(t))\right\}, \\
&&\text{where}~~V_{\text{dyonium}}(d)= K d - \frac{C}{d} +  \frac{C}{\delta}.
\end{eqnarray} 

The action of dyonium obtained in this way seems very suggestive; the classical solution of electric and magnetic fields by solving the London equation gives the Coulomb interaction between dyon and anti-dyon.   Missing in the London equation but definitely existing as Z-string of connecting dyon and anti-dyon, provides a linear potential.  Thus, the dyonium has the same type of potential for mesons, linear plus Coulomb type.

\subsection{Classical motion of length $d(t)$ of dyonium }
We have obtained the action of the dyonium in terms of the distance $d(t)$, which gives the classical motion of $d(t)$.  The energy conservation reads,
\begin{eqnarray}
\frac{M}{2}  (\dot{d})^2- \frac{C}{d} + K d =E_{\text{tot}},
\end{eqnarray}
from which, we have, (by changing the notation from $d$ to $r$, in order not to be mislead)
\begin{eqnarray}
dt = dr \sqrt{\frac{M/2}{E_{\text{tot}}+\frac{C}{r} - Kr}}.
\end{eqnarray}
This implies a periodic motion between $-d_* <d(=r) < d_*$, with a period $T$ for $-T/2 < t <T/2$.
 Here, $T$ and $d_*$ are given by
\begin{eqnarray}
\frac{T}{2}= \int_{\delta}^{d_*} d r \sqrt{\frac{M/4}{E_{\text{tot}}+\frac{C}{r} - Kr}},  ~\text{with}~ V_{\text{dyonium}}(d_*)=-\frac{C}{d_*} + Kd_*=E_{\text{tot}}.
\end{eqnarray}

Accordingly, the Chern number increases a constant amount $|q_{1/2}|$ during the half-period $T/2$,  
\begin{eqnarray}
|q_{1/2}|=\int_0^{\frac{T}{2}} dt \left(\frac{1}{\delta}-\frac{1}{d(t)} \right) =  \int_{\delta}^{r_*} dr \left(\frac{1}{\delta}-\frac{1}{r} \right)\sqrt{\frac{M/4}{E_{\text{tot}}+\frac{C}{r} - Kr}}.
\end{eqnarray}
During the multiple half-periods from $t=0$ to $t=nT/2+ \Delta t$, where $n$ (positive integer) stands for the number of repetition of half-periods, and $0< \Delta t < T/2$,  the ``Chern number of dyonium increases \underline{monotonously}, namely almost linearly with periodic modulations'',
\begin{eqnarray}
q(t)= n |q_{1/2}| + \Delta q, ~~(0< \Delta q< |q_{1/2}|, ~n=0, 1, 2, \cdots).
\end{eqnarray}

Since the Hamiltonian of dyonium is conserved as $E_{tot}$, its action also increases linearly in time,
\begin{eqnarray}
S_{\text{dyonium}}=-E_{\text{tot}}\int dt.
\end{eqnarray}

\subsection{Transition amplitude of L- and B-number violation processes}

If Chern number $q$ is generated, the B- and L-number violation occurs, as shown in Subsection 3.1.  That is, the $q-$point function of the chiral fermions can be evaluated as follows:
\begin{eqnarray}
&&\langle \psi_{L}(x_1)_1 \cdots \psi_{L}(x_n)_q \rangle \nonumber \\
&&\equiv \frac{\int \mathcal{D} A(x) \mathcal{D} B(x) \int \mathcal{D} \bar{\psi}(x)  \mathcal{D} \psi(x) \;  e^{iS_{\text{dyonium}}^{(q)}} \; e^{ i S_{\text{fermion}} }  \left(\psi(x_1)_{1} \cdots \psi(x_q)_{q} \right)}{\int \mathcal{D} A(x) \mathcal{D} B(x)  \int \mathcal{D} \bar{\psi}(x)  \mathcal{D} \psi(x) \;  e^{iS_{\text{dyonium}}^{(q)}} \; e^{i S_{\text{fermion}} }} \\
&&= \frac{\int \mathcal{D} A(x) \mathcal{D} B(x)  e^{iS_{\text{dyonium}}^{(q)}}   \left(\chi(x_1)_{1} \cdots \chi(x_q)_{q} \right)}{\int \mathcal{D} A(x) \mathcal{D} B(x)   e^{iS_{\text{dyonium}}^{(q)}} } \equiv \langle \chi(x_1)_1 \cdots \chi(x_q)_q \rangle,  \label{fermion amplitude}
\end{eqnarray}
where the zero mode $\chi(x)_i$ for the L-handed fermion $\psi_{L}(x)_i$, satisfying $S_{\text{fermion}}(\chi_i)=0$, or
\begin{eqnarray}
i \gamma^{\mu} D_{\mu}(A, B) \chi_i(x; A(x), B(x))=0,
\end{eqnarray}
gives the dominant contribution. The zero mode is obtained for each fermion species $i$, depending on a given configuration of gauge fields.

Here, we consider the case of doublet fermions.
The fermion number violation process to occur depends on the model chosen.  

\subsubsection{Simple Examples} 
The simplest example is given in the standard model (SM) with four fermion doublets, which is relevant to the proton decay.  We have four doubles of quarks and leptons in a generation,
\begin{eqnarray}
\text{SM}:  \left\{ \begin{pmatrix} \nu_e \\e \end{pmatrix}_L,
~\begin{pmatrix} u_a \\d_a \end{pmatrix}_L \right\}_{SM} ~(\text{with three colors}~a=1-3).
\end{eqnarray}

As was estimated in Eq.(\ref{fermion amplitude}), Whenever the dyonium gains Chern number $q$, the non-vanishing fermionic amplitude in the dyonium background requires the $q$ fermion zero modes, one from each doublet, 
\begin{eqnarray}
\langle \psi_{L}(x_1)_1 \cdots \psi_{L}(x_q)_q \rangle
= \langle \chi(x_1)_1 \cdots \chi(x_n)_q \rangle.
\end{eqnarray}
The amplitude should be invariant under the global symmetry of $SU(2)_L \times U(1)_Y$.

In the background field of the dyonium,  the zero mode solution plays the role of wave function for  fermion.  Following the fermion wave functions of annihilation of particle $u(\bm{p}, s)$, and creation of anti-particle $v(\bm{p}, s)$, we define 
the wave functions in the momentum representation of a fermion species $i$, in the background field of the dyonium $(D\bar{D})$,
\begin{eqnarray}
u^{(D\bar{D})}_i(p, s) \equiv \int d^4 x \; e^{i px} [\chi_i(x)]_{s_3=s}, ~~\text{and}~~v^{(D\bar{D})}_i(p, s)= u(p. s)^{C} = (i \sigma_2) u^{(D\bar{D})}_i(p. s)^*,
\end{eqnarray}
where $C$ is the charge conjugation operation.
Therefore, we can write down the fermion operator near the dyonium as follows:
\begin{eqnarray}
\hat{\psi}_i(x)_L= \sum_{s=\pm \frac{1}{2}} \int  \frac{d^3 p}{(2\pi)^3 2|\bm{p}|} \left( \hat{b}_i(p, s))  u^{(D\bar{D})}_i(p, s) e^{-ipx} + \hat{d}_i(p, s)^{\dagger}  v^{(D\bar{D})}_i(p, s) e^{ipx} \right),
\end{eqnarray}
where $\hat{b}$ and $\hat{d}^{\dagger}$ are annihilation operator of fermion $i$, and creation operator of its anti-fermion $\bar{i}$, respectively.

Then, the amplitude $\mathcal{A}\left(  u_{1L} + u_{2L} \to   \overline{d_3}_R  + e^{+}_R  \right)$, contributing to the proton decay $p \to \pi^0 + e^{+}$,  can be estimated following the ordinary field theory, by using the fermion zero modes as wave functions. 

In the same way, we can examine another example beyond the SM.  

To generate the neutrino mass, a $SO(10)$ GUT model with four fermion doublets, $\{\psi^{(0)}, \cdots, \psi^{(3)}\}$ in (\ref{eq4}), was studied in \cite{Sugamoto 1}.  If we choose $SU(2)_{ij}$ group with $(ij, klm)=(14, 235)$, then we have the following four fermion doublets:
\begin{eqnarray}
&&\text{Beyond SM}:  \begin{pmatrix} \overline{N_R} \\ u_1 \;  \end{pmatrix}_L,\; \begin{pmatrix}  \overline{u_1} \\ \nu \;  \end{pmatrix}_L, \; \begin{pmatrix}  d_3 \\ \overline{d_2} \;  \end{pmatrix}_L,  \; \begin{pmatrix}  d_2 \\ \overline{d_3} \;  \end{pmatrix}_L .
\end{eqnarray} 

Similarly, we can estimate the amplitude
$\mathcal{A}\left( N_R \to \nu_L  +  \overline{(d_c)}_L + (d_c)_L  \right) ~(c=1, 2)$, which is relevant to the generation of neutrino mass.

Now, we are ready to examine the estimation of fermion zero modes. 

\section{Fermon zero modes}
In order to estimate the fermion zero modes, we begin with the Dirac equation.
\subsection{Dirac equation}
Choosing the chiral basis \footnote{~~Chiral basis, $\gamma^0=\begin{pmatrix} 0 & I \\ I & 0 \end{pmatrix}, \;  \gamma^5=\begin{pmatrix} I & 0 \\ 0 & -I \end{pmatrix}, \; \bm{\gamma}= \begin{pmatrix} 0 & -\bm{\sigma} \\ \bm{\sigma} & 0 \end{pmatrix}, ~\text{and} ~~\psi(x)=\begin{pmatrix} \psi_R(x) \\ \psi_L(x) \end{pmatrix}. $}, the L-handed fermionic zero mode satisfies 
\begin{eqnarray}
\left\{ \left( i \partial_t +  \frac{\tau^a}{2} g A^a_{0}(x) +  \frac{Y}{2} g' B_{0}(x)\right) 
 + \bm{\sigma} \cdot \left(\bm{p} +  \frac{\tau^a}{2} g \bm{A}^a(x) +  \frac{Y}{2} g' \bm{B}(x)\right) \right\} \psi_L(x)=0.
\end{eqnarray}
We may write it with the notation $\bar{\sigma}^{\mu}$ and the   Minkowski metric,\footnote{~~The four component notation $\bar{\sigma}^{\mu}=(1, -\bm{\sigma})$, with $\bm{p}=-i \bm{\nabla}, \bm{A}^a=-A^{a}_{i}, \bm{B}=-B_{i}~\text{for}~(i=1-3).$}
\begin{eqnarray}
 i \bar{\sigma}^{\mu} D_{\mu}(A, B) \psi_L(x)=\bar{\sigma}^{\mu} \left( i \partial_{\mu} +  \frac{\tau^a}{2} g A^a_{\mu}(x) +  \frac{Y}{2} g' B_{\mu}(x)\right)  \psi_L(x)=0.
\end{eqnarray}  

Here, $Y$ is the hypercharge for the fermion species $\psi_L$.

A direct way to solve the zero mode equations gives a coupled partial differential equations among four wave functions.  Rather than this direct way, we will adopt the simpler way, based on some ansatz.  

Our way is to search for the solution as the tensor product of a spinor in the iso-space times that in the spin-state.
Furthermore, we identify the iso-spinor to the Higgs doublet $\phi(x)$.  We denote the other spinor in the spin-space as $\psi(x)$ which is so determined as the tensor product may satisfy the original Dirac equation for zero modes.  

Thus, the ansatz for $\psi_L(x)$ can be written as
\begin{eqnarray}
\psi_L(x) = e^{im\varphi} \phi(x) \otimes \psi(x)= e^{im\varphi} \begin{pmatrix} \phi(x) \psi_1(x) \\ \phi(x) \psi_2(x) \end{pmatrix}, ~~\psi(x)=\begin{pmatrix}  \psi_1(x) \\ \psi_2(x) \end{pmatrix}
\end{eqnarray}
  
Then, we starts to solve the Dirac equation with hypercharge $Y$ and $J_3=m$ in the dyonium background field, namely,
\begin{eqnarray}
&&0=\bar{\sigma}^{\mu}  i D_{\mu}(A, B)^{(Y)} e^{im\varphi}\phi(x) \otimes \psi_Y(x) \\
&&= \bar{\sigma}^{\mu}  \left(i D_{\mu}(A, B)^{(Y=-1)} + \frac{1+Y}{2}g' B_{\mu} + \sqrt{g^2 +g'^2} \frac{\tau^a}{2} A^a_{\mu}(\text{Z-string}) \right) e^{im\varphi}\phi(x) \otimes \psi_Y(x) \nonumber \\
&&=e^{im\varphi} \left( g' \frac{1+Y}{2}B_{\mu} + \sqrt{g^2 +g'^2} \frac{\tau^a}{2} A^a_{\mu}(\text{Z-string}) \right) \phi(x) \otimes \bar{\sigma}^{\mu} \psi_Y(x)   \nonumber \\
&& + \; e^{im\varphi} \phi(x) \otimes i \bar{\sigma}^{\mu}  (\partial_{\mu} +im \partial_{\mu} \varphi ) \psi_Y(x),  
\end{eqnarray}
where the London equation, $D_{\mu}(A, B)^{(Y=-1)} \phi(x)=0$ was used.

To estimate this equation, we need the explicit expressions for $B_{\mu}$ and $A^a_{\mu}(\text{Z-string})$.  The latter is the additional contribution from the $Z$-string given in (\ref{gauge field for Z}).

 Non-vanishing components for $B_{\mu}$ and $A^a_{\mu}(\text{Z-string})$ are 
\begin{eqnarray}
&&g' B_0= Q\rho(x)=-\frac{g^2}{2}\rho(x), ~g'B_{\varphi}= \eta \frac{1- \cos \Theta}{r}, \\
&& A^a_{\varphi}(\text{Z-string})= n_1^a  \frac{1}{\sqrt{g^2+g'^2}}\left(  \frac{1- \cos \Theta}{r} - 2 m_Z K_1(m_Z r) \right).
\end{eqnarray}

Here, we meet with a welcome formula, which implies that the non-Abelian isospin structure in $A^a$(Z-string)) can be completely absorbed into the Higgs doublet itself, that is,
\begin{eqnarray} 
\frac{\tau^a n_1^a}{2}  \phi(x)= \frac{1}{2} H(r)\begin{pmatrix} \cos \Theta, & \sin \Theta e^{-i\varphi} \\ \sin \Theta e^{i\varphi},   &  -\cos \Theta \end{pmatrix} \begin{pmatrix} \cos \frac{\Theta}{2} \\ \sin \frac{\Theta}{2} e^{i\varphi} \end{pmatrix}= \frac{1}{2}\phi(x).
\end{eqnarray}

Owing to this formula, the iso-doublet field $\phi(x)$ decouples from the equation, leading to a simple two component Dirac equation of iso-singlet field $\psi_Y(x)$,
\begin{eqnarray}
 \left\{ i \partial_t  + \left(\frac{1+Y}{2}\frac{-g^2}{2}\rho \right) -i \bm{\sigma} \cdot \bm{\nabla} + \sigma_{\varphi} C_{\varphi}(x) \right\} \psi_Y(x) \times f_Y(x)=0,~~\label{Eq for iso-singlet spinor}
\end{eqnarray} 
where the decomposition of $\bar{\sigma}^{\mu}$ and $\bm{A}(x)$ into components\footnote{~The notation $\sigma_{\varphi}=i(-\sigma_+e^{-i \varphi} + \sigma_- e^{i \varphi} )$, is given by
$\bm{\sigma} \cdot \bm{\nabla}=\sigma_z \partial_z + \sigma_r \partial_r + \sigma_{\varphi} \frac{1}{r}\partial_{\varphi} =\sigma_z \partial_z + (\sigma_+e^{-i \varphi} + \sigma_- e^{i \varphi})\partial_r + i(-\sigma_+e^{-i \varphi} + \sigma_- e^{i \varphi} )\frac{1}{r}\partial_{\varphi}$, 
and
$\bm{\sigma} \cdot \bm{A}= \sigma_z A_z + \sigma_r A_r + \sigma_{\varphi} A_{\varphi}.$} are used with the definition of $\sigma_{\varphi}$.

Then, the $\varphi$ component gauge field $C_{\varphi}(x)$ is given by
\begin{eqnarray}
C_{\varphi}(x)= \frac{1+Y}{2}\eta \frac{1- \cos \Theta}{r} +    \frac{1}{2\sqrt{g^2+g'^2}}\left(  \frac{1- \cos \Theta}{r} - 2 m_Z K_1(m_Z r) \right) + \frac{m}{r}. \label{C-phi}
\end{eqnarray}

What we are going to do in the following is to choose $\psi_Y(x)$, two component real-spinor, as a solution of another ``London equation'',
 \begin{eqnarray}
 \left\{ i \partial_{\mu} -  C_{\mu}(x)\right\} \psi_Y(x)=0, \label{London 2}
 \end{eqnarray}
where $C_{\varphi}(x)$ is the only non-vanishing component for $C_{\mu}(x)$.
 
\subsection{Axial symmetry}
Our dyonium solution has an axial symmetry about a rotation around the $z$-axis.  It is easily seen that if we perform the following two rotations successively, the iso-space rotation around the iso-spin's third axis and the real-space rotation around its third axis $z$ with the same angles, then the Dirac equation is invariant.  The space rotation is associated with the rotation of spin for fermions.  Therefore, the wave function undergoes a phase change by the following rotation, 
\begin{eqnarray}
e^{i \left(\hat{L}^3 +  \frac{\tau^3}{2} + \frac{\sigma^3}{2} \right)} \psi_L(x)=e^{i m \phi} \psi_L(x),
\end{eqnarray}
where $\hat{L}^3$ is the third component of the angular momentum operator.  Thus, the third component $J^3$ is conserved for the axial symmetric dyonium,
\begin{eqnarray}
J^3 = L^3 + \frac{\tau^3}{2} + \frac{\sigma^3}{2}=m.
\end{eqnarray}

If we label the wave function as $\psi_{L_3, \frac{\tau_3}{2}, \frac{\sigma_3}{2}}$, with the third components of angular momentum, iso-spin and spin, then there are four fermion states with $J_3=m$, 
\begin{eqnarray}
\psi_L(J_3=m)=\left(\psi_{m-1, \frac{1}{2}, \frac{1}{2}}, \psi_{m, -\frac{1}{2}, \frac{1}{2}}, \psi_{m, \frac{1}{2}, -\frac{1}{2}}, \psi_{m+1, -\frac{1}{2}, -\frac{1}{2}} \right)_L,
\end{eqnarray}
where all four components have the same $J^3=m$.

Therefore, the ansatz of the tensor product for the fermion zero mode, consistent with the axial symmetry, can be 
\begin{eqnarray}
\chi_L(x)= e^{im\varphi} \phi \otimes \psi_Y= e^{im\varphi} \times \begin{pmatrix} \cos \Theta \\ e^{i \varphi} \sin \Theta \end{pmatrix} \otimes \begin{pmatrix} -e^{-i \varphi} \sin \Psi_Y \\  \cos \Psi_Y \end{pmatrix} \times f_Y(x).
\end{eqnarray}
Here, we have used the normalized 2-component spinors for both $\phi$ and $\psi_Y$, so that the extra $f_Y(x)$ field is multiplied.  We have denoted the zero mode fermion as $\chi_L(x)$ as before.

\subsection{Reduction of Dirac equation to a partial differential equation for a single function $f_Y(t, \bm{x})$}

Before determining the spinor $\psi_Y(x)$ in the spin-space, let us consider why the form of gauge potentials (appeared in the dyonium solution) has a special form.

In other words, what is the origin of the form of gauge potential, $A_{\mu} \sim \frac{1-\cos \Theta}{r}\partial_{\mu}\varphi$ ?  Indeed this form frequently appears in the dyonium solution.  We know already that it comes from the spinor relation in the iso-space,
\begin{eqnarray}
&&-i \left(\phi(x)^{\dagger} \overleftrightarrow{\partial_{\mu}} \phi(x) \right)
= -i \begin{pmatrix} \cos \frac{\Theta}{2}, & e^{i \varphi} \sin \frac{\Theta}{2}  \end{pmatrix}^{\dagger} \overleftrightarrow{\partial_{\mu}} \begin{pmatrix} \cos \frac{\Theta}{2} \\ e^{i \varphi} \sin \frac{\Theta}{2} \end{pmatrix} \\
&&= 2 \sin^2 \frac{\Theta}{2} \; \partial_{\mu} \varphi= \frac{1- \cos \Theta}{r} \hat{1}_{\varphi}.
\end{eqnarray}
In the same manner, for the spinor in the ``spin-space'', if it takes the following form, 
\begin{eqnarray}
\psi_Y(x)= \begin{pmatrix} -e^{-i \varphi} \sin \frac{\Psi_Y}{2} \\ \cos \frac{\Psi_Y}{2}  \end{pmatrix},
\end{eqnarray}
we have the same relation
\begin{eqnarray}
i \left(\psi_Y(x)^{\dagger} \overleftrightarrow{\partial_{\mu}} \psi_Y(x) \right)
= 2 \sin^2 \frac{\Psi_Y}{2} \; \partial_{\mu} \varphi= \frac{1- \cos \Psi_Y}{r} \hat{1}_{\varphi}.  \label{129}
\end{eqnarray}
On the other hand, this solution of the gauge field is obtained from the London equation (\ref{London 2}).  Therefore, we have
\begin{eqnarray}
 C_{\mu}(x)= \frac{i}{2} \left(\psi_Y^{\dagger}(x) \overleftrightarrow{\partial_{\mu}} \psi_Y(x)\right), ~~\text{or}~~C_{\varphi}(x)= \frac{1- \cos \Psi_Y}{2r} .
 \end{eqnarray}

To fix $\Psi_Y(r, z)$, $C_{\varphi}(x)$ should be chosen as  Eq.(\ref{C-phi}), which implies
 \begin{eqnarray}
 1- \cos \Psi_Y \equiv  \left((1+Y) \eta + \frac{1}{\sqrt{g^2 + g^{'2}}} \right) (1- \cos \Theta ) + \frac{2 m_Z r}{\sqrt{g^2 + g^{'2}}} K_1(m_Z r)  + 2m. ~~ \label{relation}
\end{eqnarray}
Here, we have to choose a proper angular momentum $J_3=m$ to have a solution of $\psi_Y(r, z)$ for a given hypercharge $Y$. 

Now, we have determined the time-independent spinor in the spin space $\psi_Y(r, z)$, explicitly.  

Next, we are going to determine the remaining part of the wave function, that is, the time-dependent $f_Y(x)$, so that $\chi_L$ may satisfies the zero mode Dirac equation.  

Indeed, Eq.(\ref{Eq for iso-singlet spinor}) gives the following equation for $f_Y(x)$, 
\begin{eqnarray}
\psi_Y \left( i \partial_t + \frac{1+Y}{2}\frac{-g^2}{2} \rho(x) \right) f_Y - i (\bm{\sigma} \psi_Y) \cdot \bm{\nabla} f_Y =0,
\end{eqnarray}

Multiplying $\psi_Y^{\dagger}$ from the left, we have
\begin{eqnarray}
\left( i \partial_t + \frac{1+Y}{2}\frac{-g^2}{2} \rho(x) \right) f_Y - i \left(\hat{\bm{n}}_Y(x) \cdot \bm{\nabla}\right) f_Y =0, \label{eq for f}
\end{eqnarray}
where $\hat{\bm{n}}_Y(x)$ is the the vector field in the real-space, an analog of the Higgs triplet vector $n_1^a=(\phi^{\dagger} \tau^a \phi)$ in the iso-space:
\begin{eqnarray}
\hat{\bm{n}}_Y(x)=(\psi_L^{\dagger} \bm{\sigma} \psi_L)=-(\sin \Psi_L \cos \varphi, \sin \Psi_L \sin \varphi, \cos \Psi_L)_{\text{Cartesian real space}}.
\end{eqnarray}

The obtained equation Eq.(\ref{eq for f}), can be solved without much difficulty, since all the coefficient functions are explicitly given by $r$ and $\cos \Theta$.  

What we have done in the above can be summarized as follows: Complexities originally existing associated with iso-spin and real-spin structures, have been completely solved, by introducing a tensor product of iso-spinor $\phi$ and real-spinor $\psi_Y$.  These two spinors can be determined by the dyonium solution, and the remaining single component function $f_Y(t, \bm{x})$ satisfies a simple equation, which describes the dynamics of fermions.

The final equation obtained for $f_Y(x)$, is a kind of renormalization group (RG) equation for a zero-point function, in which $t$ plays a role of the logarithm of energy scale, and three parameters $(d, r, z)$ play the roles of three coupling constants.

\subsection{Renormalization group-like equation for fermion zero modes}
Let us discuss a little on the obtained renormalization group (RG)-like equation.  

From the definition of $\hat{\bm{n}}_Y$ we have\footnote{~$\nabla=\left( \cos \varphi \partial_r - \sin \varphi \frac{1}{r}\partial_{\varphi}, \sin \varphi \partial_r + \cos \varphi \frac{1}{r}\partial_{\varphi}, \partial_z\right)_{\text{Cartesian real space}}$} 
\begin{eqnarray}
\hat{\bm{n}}_Y \cdot \bm{\nabla}= - (\sin \Psi_Y \partial_r + \cos \Psi_Y \partial_z). 
\end{eqnarray}
Since the time variation is induced by the temporal change of the distance $d(t)$ between dyon and anti-dyon, the equation for $f_Y$ yields, by using the three variables $x=(d, r, z)$, as
\begin{eqnarray}
\left( \partial_t + \dot{d}(t) \; \partial_d + \sin \Psi_Y(x) \; \partial_r + \cos \Psi_Y(x) \;  \partial_z + (-i)\frac{1+Y}{2}\frac{-g^2}{2} \rho(x)  \right)  f_Y(x) =0, 
\end{eqnarray}
where the coefficient functions $\rho(x)$ and $\Psi_Y(x)$ as well as $\Theta(x)$, are given by
\begin{eqnarray}
&&\circ \; \rho(x)= \frac{1}{\sqrt{r^2 + (z- d/2)^2}} - \frac{1}{\sqrt{r^2 + (z + d/2)^2}}, \\
&&\circ \;  1- \cos \Psi_Y \equiv  \left((1+Y) \eta + \frac{1}{\sqrt{g^2 + g^{'2}}} \right) (1- \cos \Theta ) + \frac{2 (m_Z r) K_1(m_Z r)}{\sqrt{g^2 + g^{'2}}}  + 2m, ~~~\\
&&\circ \;  \cos \Theta(x)-1= \frac{z-d/2}{\sqrt{r^2 + (z - d/2)^2}}-\frac{z+d/2}{\sqrt{r^2 + (z + d/2)^2}}.
 \end{eqnarray}
 The temporal change of $\dot{d}(t)$ is classically determined as 
 \begin{eqnarray}
 \dot{d}(t)= v_d(x)= \sqrt{ 2E_{tot}/M +2C/(Md) -2Kd/M},
 \end{eqnarray}
where $C= 8\pi/e^2 \left( \sin^4 \theta_W-g^4/4 \right)$, and $K=v^2/4\pi$, and $M=g^4/2e^2 \int  dr dz \; 2\pi r f(r, z)^2$.  The equation is simply
\begin{eqnarray}
D_t \; f_Y(x, t)= \left\{ \partial_t + \sum_{i=d, r, z} v_i(x) \partial_i \right\} f_Y(x, t)= - i \gamma(x)  f_Y(x, t),
\end{eqnarray}
where $D_t$ is the Lagrangian derivative in hydrodynamics.
Then, this equation can be solved, by using the running variables $\bar{x}_i(t)~(i=d, r, z)$, denoted usually with bars,
\begin{eqnarray}
\begin{cases}
~\circ \; \frac{d(\bar{d}(t))}{dt}= v_d(\bar{x})= \sqrt{ 2E_{tot}/M +2C/(M\bar{d}) -2K\bar{d}/M}, \\
~\circ \; \frac{\bar{d}\bar{r}(t)}{dt}=v_r(\bar{x})= \sin \Psi_Y(\bar{x}), \\
~\circ \; \frac{d\bar{z}(t)}{dt}= v_z(\bar{x})= \cos \Psi_Y(\bar{x}) \\
~~~~~~~~~=1-   \left((1+Y) \eta + \frac{1}{\sqrt{g^2 + g^{'2}}} \right) (1- \cos \Theta(\bar{d}, \bar{r}, \bar{z}) - \frac{2 (m_Z \bar{r}) K_1(m_Z \bar{r})}{\sqrt{g^2 + g^{'2}}}  - 2m . 
 \label{stream line}
\end{cases}
\end{eqnarray}
with the ``anomalous dimension''
\begin{eqnarray}
\gamma(x)=\frac{1+Y}{2}\frac{g^2}{2} \rho(x).
\end{eqnarray}

Now, we can write the solution of $f_Y(x)$ formally as follows:
\begin{eqnarray}
f_Y(\bm{x}, t)= e^{ -i (\frac{1+Y}{2}\frac{g^2}{2} )\int_0^{t} dt' \; \rho(\bar{x}(t')) } \times f_Y(\bm{x}_0, t=0).
\end{eqnarray}
Here, $\bm{x}=\bar{\bm{x}}(t)$ expresses the final port at time $t$, after leaving $\bm{x}_0=\bar{\bm{x}}(0)$ at $t=0$, by taking a boat which floats with the flow of the stream.

The velocity field $\bm{v}(x)$ (or a set of $\beta$ functions) in our problem is explicitly known, so that it is expected to guide us to any place at any time $(\bm{x}, t)$, starting from a boundary point.  The boundary are the spacial infinity, $|\bm{x}| \to \infty$, and the axis of the Z-string, $r \to 0$ and $-d/2 \le z \le d/2$.   From the normalizability of the wave function, we expect the boundary condition is that $f_Y$ vanishes exponentially at the spacial infinity, while on the Z-string it approaches to zero power likely.  Anyway the detailed analysis on the boundary condition is necessary.

If everything works well, then the fermion zero mode with hypercharge $Y$ can be determined as follows:
\begin{eqnarray}
\chi_L(x)^{(Y)}= e^{im\varphi} \times \phi(x) \otimes \psi_Y(x) \times f_Y(x),
\end{eqnarray}
where $f_Y(x)$ is the solution of the RG-like equation, and two spinors $\phi(x)$ and $\psi_Y(x)$ in the iso-space and the real-space, respectively, are given by the dyonium configuration,
\begin{eqnarray}
\phi(x)=\begin{pmatrix} \cos \Theta \\ e^{i \varphi} \sin \Theta \end{pmatrix}, ~\text{and}~~
\psi_Y(x)= \begin{pmatrix} -e^{-i \varphi} \sin \frac{\Psi_Y}{2} \\ \cos \frac{\Psi_Y}{2}  \end{pmatrix}.
\end{eqnarray}
Here, $\Theta(r, z)$ and $\Psi_Y(r, z)$ are explicitly given in terms of $(d, r, z)$, if a proper angular momentum $m$ is chosen for a given $Y$.

\section{Conclusion and Discussion}

\subsection{Conclusion}

This paper studied dynonium induced fermion number violation mechanism in a $SU(2)_L \times U(1)_Y$ gauge theory with a doublet Higgs field, in which the following issues have been solved.  These issues could not be answered in the previous paper \cite{Sugamoto 1}.

1) [C. J. Goebel's question (1982)] ``How does the electric field be excited, being parallel to the dipole magnetic field of the monopolium.''  
To answer this question, we consider this time the dyonium, by generalizing the Nambu's solution of monopolium \cite{Nambu}.

\vspace{3mm}

2) The obtained dyonium oscillates, under the linear plus Coulomb potential, acting between dyon and anti-dyon. The linear potential comes from the ``$Z$-string'' connecting monopole and anti-monopole, while the Coulomb potential comes from both the electric and magnetic dipole forces.  This osicillation creates the Chern number (or the fermion number), as monotonously increasing or decreasing in time.

\vspace{3mm}

3) The gauge field configuration coming from the $Z$-string, ignored previously, is taken into account this time in the approximation of infinitely long dyonium.  The study of the Z-string for a finite sized dyonium is the next target.

\vspace{3mm}

4) Towards estimating the rate of the fermion number violation processes, we have examined the Dirac equation.  In this study we have found a way to reduce the Dirac equation of the $SU(2)_L \times U(1)$ theory to a single component renormalization group-like equation, without spin and iso-spin.  This reduction is simple, so that it is useful to evaluate the reaction rate of the fermion number violation processes explicitly as a product of the fermion zero modes.  

\subsection{Discussion}

The issues not studied well in this paper are summarized as follows:

\subsubsection{Solving RG-like equation}
[Issue 1)]  Numerical estimation of the fermion zero modes by solving the RG-like equation, and to estimate the reaction rates for the B- and L-number violation processes including the neutrino mass, are remainedl.  In this study we should consider any $SU(2)_L$ and $U(1)_Y$ groups embedded in a larger group.  It is not clear at present, but there is a hope that the RG equation can be solved easily, by separating the variables in the rotating elliptic coordinate system $(\xi, \eta)$, which will be discussed next.

\subsubsection{Rotating elliptic (ellipsoidal) coordinate system}

[Issue 2)]  A natural coordinate system for the dyonium is $(\rho, \cos \Theta, \varphi)$, since $\rho(x)$ is a potential of the dipole field, and $\cos \Theta$ depicts the equi-potential curves.  This coordinate system is, however, a curvilinear coordinate system, but has been surely studied in the past and there exist the mathematical formulae we can refer to.  Especially, we have to know the special functions on the two dimensional plane described by the coordinate system $\bm{x}_{(2D)}=(\rho, \cos \Theta)$ with fixed azimuthal angle $\varphi$:  
\begin{eqnarray}
 \left\{ \bm{\nabla}_{(2D)}^2 - \frac{m^2}{r^2} \right\} \tilde{K}_m(\bm{x}_{(2D)}, M)=M^2 \tilde{K}_m(\bm{x}_{(2D)}, M). 
\end{eqnarray}
About the expression of $\bm{\nabla}_{(2D)}^2$, see the footnote 4).

If the replacement is successfully done, from the modified Bessel function $K_m(x, M)$ to this new special function $\tilde{K}_m(\bm{x}_{(2D)}, M_Z ~\text{or} ~M_H)$, then we can solve the finite length case for $Z$-string.

In case of the infinitely long dyonium, we keep only $r$ dependency and ignore the dependency on $z$ and $d$.  For the finite sized case, $1-\cos \Theta$ seems more natural than $r$. 
Indeed, if we can impose a condition $|z| \ll d$, restricting to the central region of dyonium, then we may find a relation, 
\begin{eqnarray}
\cos \Theta \approx 1- \frac{d}{\sqrt{r^2 + (d/2)^2}}, ~~
\text{or}~~ r \approx d \sqrt{ \frac{1}{(1-\cos \Theta)^2} - \frac{1}{4} }.
\end{eqnarray}
This suggests the usage of $(1-\cos \Theta)$ instead of $r$; the former variable seems not so bad for the dyonium, since it reproduces $\cos \Theta \approx -1 $ near Z-string for $r \ll d$, and $r \approx d/(1- \cos \Theta)$ for $r \gg d/2$. The more precise discussion is, however, inevitable to elucidate the inner structure of the finite sized dyonium.

We finally recognized from the discussions with the colleagues at Open U. of Japan (OUJ) that the issue 2) can be solved completely, by using a different coordinate system, the so-called ``Rotating Elliptic Coordinate System''.  Its coordinates $(\xi, \eta, \varphi)$ are related to the cylindrical coordinates $(r, z, \varphi)$, and ours $(\rho, \cos \Theta, \varphi)$ as follows:
\begin{eqnarray}
&&r= \frac{d}{2} \sqrt{(\xi^2-1)(1-\eta^2)}, ~z=\left(\frac{d}{2}\right) \xi\eta, \\
&&\rho=\left(\frac{4}{d} \right)\frac{\eta}{\xi^2-\eta^2}, ~1- \cos \Theta = \frac{2 \xi (1- \eta^2)}{\xi^2-\eta^2},
\end{eqnarray}
where the oblate-type elliptic coordinates are chosen.
The range of variables can be understood as $1 \le \xi \le \infty$ and $-1 \le \eta \le 1$, from the other parametrization of $(\xi=\cosh u, \eta=\cos v)$. The $\xi=1$ represents the finite length $Z$-string and $\xi=\infty$ is the spacial infinity, while $\eta=\cos v$ gives the periodic coordinate along the elliptic curves with two foci located on the positions of dyon and anti-dyon.

Its Laplacian is well-known,
\begin{eqnarray}
\left(\frac{d}{2} \right)^2\nabla_{(2D)}^2=  \frac{1}{\xi^2-\eta^2} \left\{ \partial_{\xi}  (\xi^2-1) \partial_{\xi}  + \partial_{\eta}  (1-\eta^2) \partial_{\eta} - m^2 \left( \frac{1}{\xi^2-1} + \frac{1}{1-\eta^2}\right) \right\}.
\end{eqnarray}

Thus, the wave equation at rest (without momenta) with mass $M$, can be 
\begin{eqnarray}
&&  \left\{ \partial_{\xi}  (\xi^2-1) \partial_{\xi}  + \partial_{\eta}  (1-\eta^2) \partial_{\eta} -  m^2 \left( \frac{1}{\xi^2-1} + \frac{1}{1-\eta^2}\right) \right\} \psi(\xi, \eta) \nonumber \\
  &&= \left(\frac{Md}{2}\right)^2 (\xi^2-\eta^2) \psi(\xi, \eta).
\end{eqnarray}
Surprisingly, the wave function can be solved by the separation of variables.  The solution is a superposition of $\psi(\xi, \eta)= \sum_{l, m}  c_{lm} \tilde{K}_{lm} (\xi)  \tilde{Y}_{lm} (\eta)$, each factor of which satisfies
\begin{eqnarray}
&&\left\{ \partial_{\eta}  (1-\eta^2) \partial_{\eta} - \frac{m^2}{1-\eta^2} + \left(\frac{Md}{2}\right)^2 \eta^2 \right\} \tilde{Y}_{lm}( \eta)=\lambda_{lm}\tilde{Y}_{lm}( \eta), \\
&&\left\{ \partial_{\xi}  (\xi^2-1) \partial_{\xi}  -  \frac{m^2}{\xi^2-1} - \left(\frac{Md}{2}\right)^2 \xi^2 \right\} \tilde{K}_{lm}(\xi)= -  \lambda_{l, m}\tilde{K}_{lm}( \xi).
\end{eqnarray}
We used a suggestive notation which implies $\tilde{Y}_{lm}( \eta)$ is a generalization to the dipole case of the spherical harmonics for $d=0$, while $\tilde{K}_{lm}(\xi)$ is that of the modified Bessel function for the cylindrical case of $d \to \infty$.

To go a step forward by writing down $\bm{A}^{(Z)}$ in terms of $\tilde{Y}_{lm}( \eta) \times \tilde{K}_{lm}(\xi)$, we have to be familiar with the special functions in the rotating elliptic coordinates (see for example \cite{elliptic coordinate}).  So, we put this issue to the next target.

\subsubsection{Production and decay of dyonium}

[Issue 3)] How is the dyonium produced in the history of the universe.  It is interesting to note that the dyonium solution is quite similar to the meson states. Therefore, the knowledge on the quarkonium and the monopolium \cite{Neil} can be utilized.  
indeed the quantum mechanical treatment of pair creation of dyon and anti-dyon, and the decay of dyonium into hadrons or multi-photons, can be discussed in the similar manner. 

Dyoniums may be produced in the reheating era at temperature $T_R$ after the inflation ends, when the inflaton decays into a pair of $D$ and $\overline{D}$: $\sigma_{\text{prod}}(\text{inflaton} \to D, \overline{D}).$  We are able to predict the number density of dyonium $(D\overline{D})$ when the reheating ends.
Afterwards, the produced dyonium starts the periodic oscillation, during which the various fermion number violation processes are induced, with the quarks and leptons existing in the neighborhood of it.  The dyonium finally decays into multi-photons, when the dyon $D$ and the anti-dyon $\overline{D}$ collide at $d < 2 \delta$ ($\delta$ is the radius of dyon and anti-dyon), $(D\overline{D}) \to 2\gamma s, \; 3\gamma s, \; \cdots$ may occur quite frequently.

It is also expected to find the dyonium at the collider experiments. The rate of production at the collider, $\sigma_{\text{prod}}(q+\overline{q}, g + g \to D, \overline{D})$, and its decay rate $\sigma_{\text{decay}}((D\overline{D}) \to \text{multi}-\gamma s)$, as well as the life-time $\tau_{\text{dyonium}}$ of the dyonium, can be estimated, by using the quantum field theory as a generalization of monopolium to dyonium (see \cite{Neil}). 

The fermion number violation processes (of neutrino mass and of B- or L-number violation) occur within the life-time of the dyonium.

\vspace{3mm}

The author leaves these unsolved issues to the future study, if possible by himself or someone else who is interested in this topic.

\section*{Acknowledgments}

The author gives his sincere thanks to Professor Charles J. Goebel for his essential question raised to him at Madison in 1982.  The main purpose of this paper is to answer to his question.  He also thanks Kaoru Hagiwara for inviting him to give a seminar at Wisconsin U., where we had an opportunity to discuss with Professor Goebel for long hours. 

He is grateful to Neil David Barrie, Mathew Talia and Kimiko Yamashita for a number of valuable discussions and comments.  He gives his thanks to the colleagues in OUJ, Shiro Komata, So Katagiri, Yoshiki Matsuoka, Mamoru Sugamoto, Koichiro Yamaguchi and Ken Yokoyama for valuable discussions, reading the manuscript and giving useful comments.




\begin{thebibliography}{99}
\bibitem{Baryogenesis}
A. D. Sakharov, Pisma Zh. Eksp. Teor. Fiz., 5, 32, [Usp. Fiz.
Nauk161,61(1991)] (1967). 

\bibitem{Manton}
N. S. Manton, ``Topology in the Weinberg-Salam theory'', Phys. Rev. D28 2019 (1983); \\
F. R. Klinkhammer and N. S. Manton, ``A Saddle point solution in the Weinberg-Salam Theory'', Phys. Rev. D30, 2213 (1984).

\bibitem{Arnold}
P. Arnold and L. MacLerren, Phys. Rev. D36, 581 (1987);  \\
J. Ambj$\phi$rn, T. Askgaard, H. Porter, M. E. Shaposhnikov, Phys. Lett. B244, 479 (1990).

\bibitem{Yanagida}
M. Fukugita and T. Yanagida, Phys. Lett., B174, 45 (1986).

\bibitem{Yoshimura} 
M. Yoshimura, Phys. Rev. Lett. {\bf 41}, 281 (1978); Erratum {\it i.b.d.} {\bf 42}, 746 (1979). 

\bibitem{A-D} 
I. Affleck and M. Dine, Nucl. Phys. B199, 251 (1987). 

\bibitem{D-S}
S. Dimopoulos and L. Susskind, Phys. Rev. D18, 4500 (1978).

\bibitem{K-K}
A. G. Kohen and D. Kaplan, Phys. Lett. B199, 251 (1987).

\bibitem{ours 1}
A. Sugamoto, ``Baryon Asymmetry: Evidence of CP Violation and Phase Transition in the Early Universe'', in Proc. the 23rd INS symposium (1995); arXiv:hep-ph/9505342.

\bibitem{ours 2}
M. Aoki, N. Oshimo and A. Sugamoto, PTP 98 (1997) 1179.

\bibitem{ours 3}
N. D. Barrie, A. Sugamoto, T. Takeuchi and K. Yamashita, JHEP 08 (2020) 072.

\bibitem{Sugamoto 1}
A. Sugamoto, ``The neutrino mass and the monopole-antimonopole dumb-bell system in the SO(10) grand unified model'', Phys. Lett. B127 (1983) 75.
\bibitem{Rubakov}
V.A. Rubakov, Pis'ma Zh. Eksp. Teor. Fiz. 33 (1981)
658 [JETP Lett. 33 (1981) 644];  C.G. Callan, Phys. Rev. D25 (1982).2141.
\bibitem{Nambu}
Y. Nambu, Nucl. Phys. B130 (1977) 505.

\bibitem{Van Allen}
Van Allen, J. Geophys. Res. 64, 271 (1959).

\bibitem{Fermi-Rossi}
E. Fermi, ``Nuclear Physics'', Chapter X-D ``Motion of charged particles in the earth's magnetic field'', U. of Chicago Press, Chicago (1951); \\
B. Rossi and S. Gilbert, ``Introduction to the physics of space'' (1970).

\bibitem{N-N}
H. B. Nielsen and M. Ninomiya, Phys. Lett. B130, 389 (1983).

\bibitem{Schwinger}
J. Schwinger, Phys. Rev. 144, 1087 (1966).

\bibitem{G-L}
L. D. Landau and V. L. Ginzburg, JETP 20 (1950) 1064.

\bibitem{N-O}
H. Nielsen and P. Olesen, Nucl. Phys. B 61 (1973) 45.

\bibitem{Neil}
N. D. Barrie, A. Sugamoto, K. Yamashita, PTEP 2016, 113B02; \\
N. D. Barrie, A. Sugamoto, M. Talia, K. Yamashita, Nucl. Phys. B 972 (2021) 115564.

\bibitem{elliptic coordinate}
Standard references: ``Iwanami Dictionary of Mathematics (3rd Edition in Japanese)'' (1985), ed. by Mathematical Society of Japan, Iwanami-Shoten Publisher, Tokyo, Japan; its 4th Edition is available in English; \\
``Iwanami formulae of Mathematics I and III''  (in Japanese)  (1990) by S. Moriguchi, K. Udagawa and S. Hitotsumatsu,  Iwanami-Shoten Publisher, Tokyo, Japan).  The more detailed formulae seem to be neccesary.


\end{thebibliography}
\end{document}